\documentclass[]{scrarticle}
\usepackage{amsmath}
\usepackage{url}
\usepackage{multirow}
\usepackage{pdflscape}  
\usepackage{graphicx}
\usepackage{etoolbox}
\usepackage{tabularx}
\usepackage{rotating}
\usepackage{wasysym}
\usepackage[utf8]{inputenc}
\usepackage[T1]{fontenc}
\usepackage[english]{babel}
\usepackage{amsfonts}
\usepackage{amssymb}

\usepackage{graphicx}
\usepackage{xspace}
\usepackage{acro}               
\usepackage[%
colorlinks= true,           
breaklinks= true,           
bookmarks = true,           
linktoc = all               
] {hyperref}    		    
\usepackage{algorithm2e}    
\usepackage[%
hyperref,                   
dvipsnames,                 
cmyk]                       
{xcolor}                    
\usepackage[%
obeyFinal                   
]%
{todonotes}                 
\usepackage{paralist}
\usepackage{tikz}
\usetikzlibrary{positioning}
\usepackage{booktabs}           
\usepackage{multirow}           
\usepackage{multicol}           

\def\rotd{\rotatebox}
\newcommand{\rot}[1]{\rotd{85}{#1}}

\newcommand{\pp}{privacy-preserving\xspace}
\newcommand{\sota}{state-of-the-art\xspace}

\newbool{abstr}
\booltrue{abstr}
\boolfalse{abstr}

\newbool{commands}
\booltrue{commands}
\newcommand{\jas}[1]{\ifbool{commands}{\textcolor{red}{#1}\xspace}{}}

\presetkeys{todonotes}{color=notes-color, size=\tiny}{}

\definecolor{unima-comment}{rgb}{0.36, 0.54, 0.66}
\definecolor{gematik-comment}{cmyk}{0.2, 0, 1.0, 0.15}

\DeclareAcronym{he}{%
	short = {HE},
	long  = {Homomorphic Encryption}}

\DeclareAcronym{fhe}{%
	short = {FHE},
	long  = {fully Homomorphic Encryption}}

\DeclareAcronym{ml}{%
	short = {ML},
	long  = {Machine Learning}}

\DeclareAcronym{mpc}{%
	short = {MPC},
	long  = {Multi-Party Computation}}

\DeclareAcronym{idash}{%
	short = {iDASH},
	long  = {integrating Data for Analysis, Anonymization, and Sharing}}

\DeclareAcronym{nn}{%
	short = {NN},
	long  = {Neural Network},
	short-plural       = {s},
	long-plural        = {s}
	}

\DeclareAcronym{rlwe}{%
	short = {RLWE},
	long  = {Ring Learning with Error}}

\DeclareAcronym{tee}{%
	short = {TEE},
	long  = {Trusted Execution Environements}}

\DeclareAcronym{gc}{%
	short = {GC},
	long  = {garbled circuits}}

\DeclareAcronym{ss}{%
	short = {SS},
	long  = {secret sharing}}

\DeclareAcronym{ot}{%
	short = {OT},
	long  = {oblicious transfer}}

\DeclareAcronym{dh}{%
	short = {DH},
	long  = {data holder},
	short-plural       = {s},
	long-plural        = {s}}

\DeclareAcronym{dp}{%
short = {DP},
long  = {data processor},	
short-plural       = {s},
long-plural        = {s}}

\title{Report: State of the Art Solutions for Privacy Preserving Machine Learning in the Medical Context}
\author{Jasmin Zalonis\footnote{University of Mannheim, Mannheim, Germany {\texttt first.last@uni-mannheim.de}} \and Frederik Armknecht\footnotemark[1] \and Björn Grohmann\footnote{gematik GmbH, Berlin, Germany {\texttt first(ö=oe).last@gematik.de}} \and Manuel Koch\footnotemark[2]}

\begin{document}
	
	\maketitle             
	
	
	\acresetall          

	\section{Introduction} \label{sec:introduction}
\begin{figure}
	\centering
 	\includegraphics[scale=0.4]{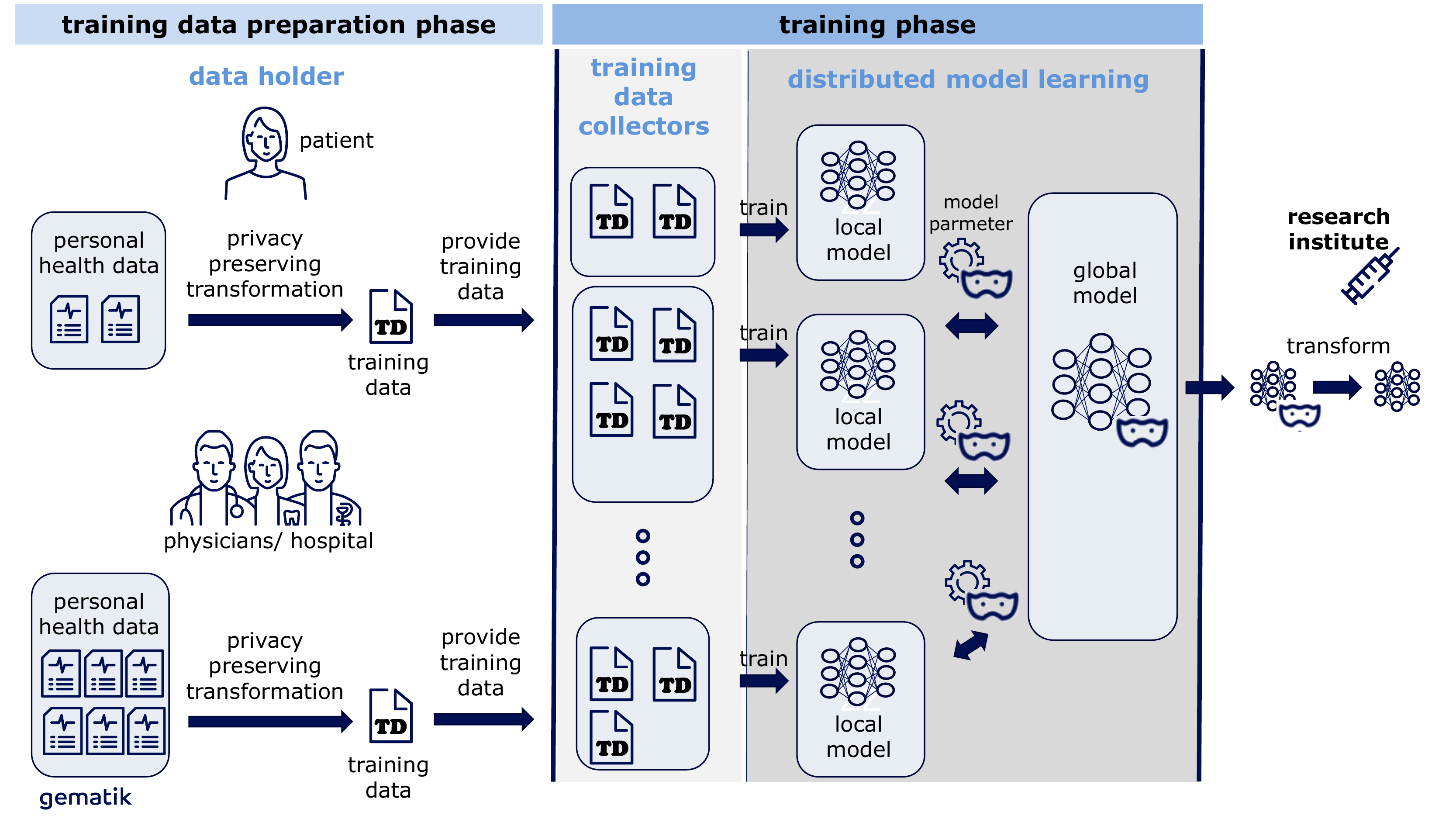}
 		\vspace{-0.5cm}
	 	\caption{Scenario of the infrastructure for training a ML algorithm.}
	 	\label{fig:trainingScenario}
\end{figure}

\begin{figure}
	\centering 
	\includegraphics[scale=0.5]{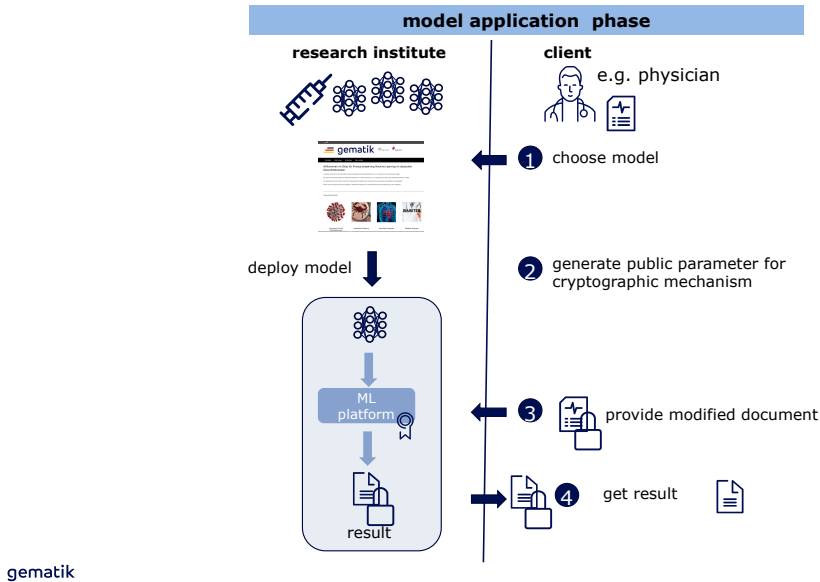}
	\vspace{-1.5cm}
	\caption{Scenario of the infrastructure for evaluating a ML algorithm.}
	\label{fig:applScenario}
\end{figure}

\setcounter{page}{1}
\ac{ml} on Big Data gets more and more attention in various fields. Over the last years a lot of improvement happened in the field of \ac{ml}. Originally, the assumed scenario was as follows: One party has a big amount of data in the clear and wants to learn something on the data. But this has changed in the last years. Privacy-preserving techniques become more important, even necessary due to legal regulations such as the General Data Protection Regulation (GDPR) and on the other hand data is often distributed among various parties. Especially in the medical context there are several \acp{dh}, e.g. hospitals or the patient itself and we need to deal with highly sensitive values. Whenever sensitive data is distributed among parties the standard scenario with one data processor that holds the data in clear is not applicable any more.  A real world scenario would be data that is held in an electronic patient record that is available in many countries by now. In Germany, for example, the "elektronische Patientenakte" (ePA) is available since 2021. It provides their users with the functionality of storing medical data that can then be used for further diagnosis or research. The medical data is encrypted. Users (e.g. physicians, hospitals) can only decrypt the data after patient authorization. One of the main questions concerning this scenario is whether it is possible to process the data for research purposes without violating the privacy of the data owner. 
We want to evaluate which cryptographic mechanism can be used in the scenario stated in Figure \ref{fig:trainingScenario}.
On the left-hand side we can see that the patient holds his own data and wants to protect his sensitive information, on the right-hand side we have the research institute that wants to gain new information from patient data by applying a \ac{ml} algorithm on a data set. In the following, we want to describe an approach where these two interests can be united. Each \ac{dh}, i.e. each patient, can choose which research institute can learn from his data, but it is crucial that nobody has access to the clear text data except the \ac{dh} itself. In the middle we have an infrastructure provider processing the patient data for model training. It is crucial, that the infrastructure provider does not learn anything about the patient. The patient sends his data either encrypted to one \ac{dp} or distributed to several \acp{dp}. After the training phase of the \ac{ml} algorithm is done the research institute gets the trained model in a secure form and can use it to gain new information or even to offer his algorithm as a service, again in a \pp way, see Figure  \ref{fig:applScenario}. 
In this context we want to explore mainly three different mechanisms: \ac{he}, \ac{mpc} and \ac{tee}. The above scenario changes a little with each technique. For \ac{he} we assume that each \ac{dh} encrypts his data with the public key of the research institute but instead of sending it directly to the research institute, which would allow them to access the data in clear, they send it to one of the \acp{dp}. We assume there are at least two \acp{dp} and the final data set is distributed among them, so no \acp{dp} has the whole encrypted data set. They both train individual a local machine learning model, eventually exchanging some intermediate data in encrypted form and therefore generating a global model, which is send back to the research institute. The institute now can decrypt the model and has only access to the learned model. 
In contrast the scenario in \ac{mpc} does not include any encryption. Rather we assume that each \ac{dh} splits his data into so called shares, pieces that alone do not reveal something about the data record. Each \ac{dp} gets one piece, one share of the data and then compute together with the other \acp{dp} the algorithm for training the \ac{ml} model. The research institute gets the model as final result as before, for example in form of all the shares which can then be recombined.  
In \ac{tee} we assume that each server (\ac{dp}) has a secure enclave. We assume full trust in this enclave. The patients as \acp{dh} send their data to a server in encrypted form. The data can only be decrypted inside the secure enclave, which trains a local \ac{ml} algorithm. Training of the global model requires communication between the enclaves, either in encrypted form, so that only the enclaves can decrypt it or in a masked form, so that the intermediate results that are exchanged do no reveal anything about the data.  
Especially in the context of \ac{mpc} the scenario where we have several clients that do not want to be involved in the computation but also do not want to provide their data in clear, is called the client-server model.
Overall we assume a semi-honest security model, that is the \acp{dp} merely try to gather information out of the protocol, but do not deviate from the protocol specification itself. In the \ac{mpc} scenario we additionally assume that we have an honest majority. 

It should be noted, that in the above scenario we do not want to clarify the question which functions, i.e. \ac{ml} algorithms are save to compute. We do not consider indirect leakage through the model parameters or the predictions of the model. It is out of the scope of this work to investigate which models are save in a way that the research institute does not learn something about the data from the generated model. Rather we want to investigate the direct leakage, e.g. how can we compute the model while minimizing avoiding the damage to privacy by addressing following questions:
Which cryptographic technique makes the most sense for the above scenario, \ac{he}, \ac{mpc}, or \ac{tee}?
 Which \ac{ml} algorithm is suited the best for the distributed scenario?
We want to address the questions first by giving a quick overview of the standard \ac{ml} algorithms and the three cryptographic mechanism. Following we will look at the \sota solutions in the recent research and with this information conclude to give a recommendation. Since the training of the algorithms are the more challenging task we focus on paper that consider the training phase and won't discuss paper who consider only the evaluation of the previous trained model.

	\section{Cryptographic Techniques} \label{sec:techniques}

\subsection{Homomorphic Encryption}
\Iac{he} is an encryption technique that allows computation on ciphertext while preserving the possibility to decrypt to the corresponding computations on the plaintexts. A distinction is made between partially \ac{he}, where only some operations can be computed in a homomorphic way, leveled or somewhat \ac{he}, where only a limited number of computations can be correctly evaluated and \ac{fhe} where arbitrary unlimited computations on encrypted data can be evaluated. The security of the most \ac{fhe} schemes is based on the hardness of the \ac{rlwe} problem.
Where a standard encryption scheme has three algorithms key-generation $Gen$, encryption $Enc$ and decryption $Dec$, a homomorphic encryption scheme has an additional algorithm evaluation $Eval$ that takes an supported function f, a set of ciphertexts and computes the corresponding function on the ciphertexts. Additionally the scheme needs to expand the correctness to the evaluation, i.e. $Dec(Eval(f, Enc(m_1), ...,  Enc(m_n))) = f(m_1, ..., m_n)$ and needs to fulfill a compactness property, meaning that the size of the output of $Eval$ is independent of the size of the evaluated function.
The standard scenario for \ac{he} is out-sourced computation, meaning a party has a big amount of data but not the resources to process these data. So in order to get to the result it sends the data in encrypted form to another party, which then can process the data in encrypted form and returns the encrypted result. Even when the service provider's system is compromised, the data would remain secure. The most popular schemes today are BGV \cite{Brakerski2014}, CKKS \cite{HEAAN}, BFV \cite{Fan2012} and TFHE \cite{TFHE}, supported by various libraries, e.g HElib, PALISADE, Microsoft SEAL and TFHE.
\ac{fhe} is a promising technique to enable a untrusted server to perform analyses on data. Because all input and output data as well as all intermediate results are encrypted privacy is preserved. With solutions that use only \ac{fhe} we need to trust that the server performs the algorithm as intended.

\subsection{Multi-Party Computation}
In contrast the goal of \ac{mpc} is for two or more parties to jointly compute a function over their inputs while keeping those inputs private and hidden from the other party. Informally the basic properties that a \ac{mpc} protocol must fulfill are: \textit{input privacy}, no information about the private data can be inferred from the sent messages than the information that can be inferred from seeing the output of the computed function alone; \textit{Correctness}, the parties obtain the correct output, even when some parties misbehave. Sometimes also \textit{fairness} is seen as a property of \ac{mpc} protocols, meaning that if one party learns the output, then all of the parties learn the output.  
Especially \ac{mpc} protocols with more than two parties often rely on \ac{ss}, where a secret input is shared among a number of parties by distributing a share of the input to each party, as described in Section \ref{sec:introduction}. Shamir \ac{ss} and additive \ac{ss} are the two commonly used protocols. Other protocols like \ac{gc} are based on two or more server, which have their own private data. Since this is not fit for our use case we will only look at paper, that accomplish our requirement of a distributed data set among the servers.
There are mainly two security models in \ac{mpc}.
\begin{compactitem}

	\item[-] Semi-Honest or Passive Security: Corrupted parties only cooperate to gather information but do not deviate from the protocol specification. Also called honest but curious adversary.
	\item[-] Malicious or Active Security: In this scenario the adversary parties are also allowed to deviate from the protocol. Therefore protocols that achieve security in an active setting yield in a higher security guarantee by detecting malicious behavior. When it comes to handling the malicious behavior the protocol can either abort and no one receives the output, or can guarantee robustness/fairness where everybody receives the correct output. The last variant is the most expensive variant.
\end{compactitem}

\subsection{Trusted Execution Environments}
\ac{tee} also called secure enclaves, as for example Intel SGX \cite{IntelSGX} and ARM TrustZone \cite{ARMTrustZone}, enable execution of programs in secure hardware enclaves in an untrusted system. In our scenario we assume we have an application that wants to train or evaluate a \ac{ml} algorithm. The clients send their sensitive data in encrypted form to the application, which can do computations with the help of the enclave on these secrets, without ever having access to the data in the clear. The confidentiality and integrity of the executed code and data are protected inside the enclaves. 
The server on which the enclave runs can seen as untrusted but we need to have trust in the enclave.

	\section{Machine Learning Algorithm}\label{sec:ML}

The field of \ac{ml} algorithm is very wide and there are lots of different algorithms and their advancements. We want to give a quick overview of the basic algorithms. In general \ac{ml} algorithm are divided into supervised and unsupervised algorithms. In supervised \ac{ml} there are both training data with an explicit label and test or evaluation data whereas in unsupervised machine there are no labeled data records. In supervised \ac{ml} the training data is used to train a model and predict the outcomes of the evaluation data on the basis of the learned model. Supervised learning is additional divided into regression models, e.g. linear regression and classification/categorization models, e.g. decision trees (classification trees) or logistic regression. Neural Networks (NN) are also supervised \ac{ml} algorithms that can be used both for regression and categorization. 
Unsupervised \ac{ml} algorithms do not use training data and therefore do not predict based on a trained model. Rather they aim to find structure in given data. Popular unsupervised algorithms are for example clustering algorithms.
Since our use case is to train a model with some ground data and later on provide the model for further prediction only supervised \ac{ml} algorithms are practicable. We will shortly describe these algorithms and how they can be trained in the following.
We assume a labeled training set $X = (x^i, y^i)^n_{i=1}$ with $n$ data records $x^i = (x^i_1, ...., X^i_m)$, $m$ features and $y$ as the corresponding labeled output. 
\subsection{Linear Regression}
The idea of Linear Regression is to fit a linear function to the input data, so that the function passes as close as possible to all of the data points. The trained model will consists of a weight-vector $\theta = (\theta_1, \theta_2, ... , \theta_m)$ and to predict the label of an evaluation record $x$ we simply compute the linear function $h_\theta(x) = \sum_{j=1}^{m}\theta_j\cdot x_j$ where $h_\theta \in \mathbb{R}$ is a continuous value. To train the model, i.e. to receive the weights that best fit the training data, we want to minimize the costs of the predicted outcome versus the real outcome (labels). The cost function can be defined with different metrics. For example the mean squared error: $J(\theta) = \frac{1}{2n}\sum_{i=1}^{n}(h_\theta( x^i) - y^i)^2$ or the L2-norm: $J(\theta) = \sqrt{\sum_{i=1}^{n}(h_\theta( x^i) - y^i)^2}$. The weights can theoretically be computed as $\theta = (X^T\cdot X)^{-1} \cdot X^T\cdot Y$ but for big data sets this has a high computational effort. There also exists iterative methods to compute $\theta$ with less effort. Because most of the paper listed in Section \ref{sec:exSolutions} use a variant of the gradient descent we will present the main idea of this iterative algorithm.
\paragraph{Gradient Descent} is an optimization algorithm to find a minimum of a function, in the case of linear regression the cost function $J_\theta$. The main idea is to start with some initial arbitrary values for the weights $\theta_1, ... , \theta_m$  and keep changing them according to the derivative to reduce $J_\theta$, see Algorithm \ref{alg:GD}. There are three big variants of the gradient descent algorithm in \ac{ml}. In the \textit{stochastic} gradient descent we will go through the data set using one record per iteration and if the function did not converge after one round in the data set we walk through the data set again. In \textit{(mini-)batch} gradient descent a batch of data is used in each iteration until convergence. When the batch is smaller than the whole data set, it is sometimes called mini-batch gradient descent, which combines the advantages of stochastic and batch gradient descent. 
\begin{algorithm}
\DontPrintSemicolon
repeat simultaneously until convergence \{ \\
\quad	$\theta_j =  \theta_j - \alpha \frac{\partial }{\partial \theta_j}J_\theta$ \quad for $j = 1, ... m$ \tcp*{with $\alpha$ as learning rate}
\}	
\caption{Gradient Descent}\label{alg:GD}
\end{algorithm}
\subsection{Logistic Regression}
Logistic Regression is a classification algorithm, where the predicted values take on only a small number of discrete values, e.g. in the binary classification problem the values $0$ and $1$ ore sometimes also $-1$ and $1$. The model again consists of a weight vector $\theta$ and the model uses the sigmoid function or sometimes called logistic function to predict a value in the range of $0$ and $1$: $h_\theta(x) = \sigma(\theta^Tx)$ with $\sigma(z) = \dfrac{1}{1 + e^{-z}}$. The sigmoid function $\sigma$ maps any real number to the $(0,1)$ interval. $h_\theta(x)$ gives the probability, that our data record belongs to the class 1. Often the classification is done based on the threshold $\tau = 0.5$, meaning that every record with $h_\theta(x) \geq 0.5$ belongs to the class 1. The cost function for logistic regression looks like 
$J(\theta) = -\dfrac{1}{n}\sum_{i=1}^{n}\left( y^i log\left( h_\theta(x^i)\right) + (1-y^i)log\left( 1 - h_\theta(x^i)\right) \right) $ and is called logistic loss or binary cross entropy. To build a model with more than two categories we basically train a logistic regression classifier for each class (throwing all the other classes into a single second class) and for a prediction pick the class that has the highest probability. The model can be trained using the gradient descent algorithm (Algorithm \ref{alg:GD}). Another popular algorithm is Newton's Method to train a logistic regression model.
\paragraph{Newton's Method} 
is an iterative method that either can be used to find the root of a differentiable function or in our context to find the root of the derivative of a function in order to find the critical points (minima, maxima or saddle points) of $f$: Given a twice differentiable function we want to solve the optimization problem $min_{x \in \mathbb{R}^n}f(x)$, see Algorithm \ref{alg:NR}. 
In the context of the logistic regression we want to minimize the cost function, therefore $f(\theta) = J(\theta)$
\begin{algorithm}
	\DontPrintSemicolon
	repeat  until convergence \{ \\
	\quad	$\theta^{t+1} =  \theta^t - H_J(\theta^t)^{-1}\triangledown J(\theta^t)$\\
	\} \\
	\tcp{\footnotesize  with $H_J(\theta)$ is the Hessian-Matrix with $H_{ij} = \dfrac{\partial^2 J(\theta))}{\partial \theta_i \partial \theta_j}$ \\
		and $\triangledown J(\theta) =\begin{pmatrix} \dfrac{\partial J(\theta)}{\partial \theta_0 } \\ ... \\ \dfrac{\partial J(\theta)}{\partial \theta_n } \end{pmatrix} $ the gradient 
	}
	\caption{Newton's Method for Optimization}\label{alg:NR}
\end{algorithm}
Where gradient descent is a simpler algorithm, newton's method does not need parameters and typically needs fewer iterations and therefore converges much faster. One downside is that newton's method has more expensive iterations and therefore is not fit for all use cases. 
\subsection{Decision Trees}
Decision trees can either be part of regression (regression tree) or a part of classification (classification tree), yet the main idea is the same. Therefore we will only describe the classification trees.
The model consists of a binary tree with several thresholds $\tau_i$ at the inner nodes. Each leaf node corresponds to one class. A new data record $x$ is classified by walking down the tree from the root to the leafs. At each inner node $i$ there is a decision criteria $\tau_i$ for one of the features of $x$ to decide on which branch to take next. 
There exists non-parametric learning techniques that can be used to produce either classification or regression trees, for example ID3, C4.5, C5.0. All of them share the same idea. At the root node start with all training examples, i.e. the hole data set $X$ and select an attribute on the basis of an splitting criteria. We then partition the instances according to the selected attribute recursively. The partitioning stops when there are no examples left or all examples for a given node belong to the same class (= purity) or there are no remaining attributes for further partitioning. When it comes to select an attribute for partitioning, we want to select the attribute, that gives the most information. A common used metrics is the Gini-Index: $Gini(D) = 1 - \sum_{i = 1}^{c} p_i^2 $ with $c$ the number of classes and $p_i$ the relative frequency of class $j$ in the data set $D$. If the data set $D$ is split on attribute $k$ into $l$ subsets the Gini-Index is defined as $Gini_k(D) = \sum_{i=1}^{l}\dfrac{|D_i|}{|D|}\cdot Gini(D_i)$. We aim to minimize the Gini-index. 
Another popular metric is the information gain, based on entropy $H(D) = -\sum_{j=1}^{c}p_j\cdot log_2(p_j)$. The information gain is the measure of the difference in entropy from before to after the set $D$ is split on an attribute $k$: $IG_k(D) = H(D) - \sum_{i=1}^lp(D_i)H(D_i)$. 
\paragraph{Random Forest }
are algorithm that can be used both for classification and regression, the algorithm builds multiple decision trees and merges the result back together by either computing the mean or average for a regression problem or returning the majority decision of all decision trees in the forest. 
\subsection{Neural Networks}
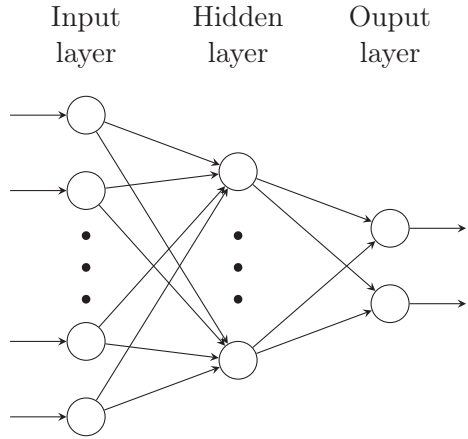
\begin{figure} 
	\begin{center}
		\tikzset{%
			every neuron/.style={
				circle,
				draw,
				minimum size=0.5cm
			},
			neuron missing/.style={
				draw=none, 
				scale=3,
				text height=0.333cm,
				execute at begin node=\color{black}$\vdots$
			},
		}
		
		\begin{tikzpicture}[x=1cm, y=1cm, >=stealth]
			
			\foreach \m/\l [count=\y] in {1,2,missing,3,4}
			\node [every neuron/.try, neuron \m/.try] (input-\m) at (0,2.5-\y) {};
			
			\foreach \m [count=\y] in {1,missing,2}
			\node [every neuron/.try, neuron \m/.try ] (hidden-\m) at (2,2-\y*1.25) {};
			
			\foreach \m [count=\y] in {1,2}
			\node [every neuron/.try, neuron \m/.try ] (output-\m) at (4,1-\y) {};
			
			\foreach \l [count=\i] in {1,2,3,4}
			\draw [<-] (input-\i) -- ++(-1,0)
			node [above, midway] {};
			
			\foreach \l [count=\i] in {1,n}
			\node [above] at (hidden-\i.north) {};
			
			\foreach \l [count=\i] in {1,n}
			\draw [->] (output-\i) -- ++(1,0)
			node [above, midway] {};
			
			\foreach \i in {1,...,4}
			\foreach \j in {1,...,2}
			\draw [->] (input-\i) -- (hidden-\j);
			
			\foreach \i in {1,...,2}
			\foreach \j in {1,...,2}
			\draw [->] (hidden-\i) -- (output-\j);
			
			\foreach \l [count=\x from 0] in {Input, Hidden, Ouput}
			\node [align=center, above] at (\x*2,2) {\l \\ layer};
			
		\end{tikzpicture}
	\end{center}

	\caption{Exemplary setup of a \acl{nn}} \label{fig:nn}
\end{figure}

A \ac{nn} is a \ac{ml} algorithm that consists of different nodes, where each node basically is a own \ac{ml} algorithm, e.g. each node could be a logistic regression. In general a \ac{nn}  is built of different layers, an input and and output layer and often in between several hidden layers. Each layer has some nodes, called neurons. The number of neurons of the output layer is defined by the desired output, e.g. for a simple binary classification problem the output layer has two nodes (see Figure \ref{fig:nn}).

 The model consists of several weight vectors $\theta^k = (\theta^k_i,j)_{i\in\{1, ..., l_k\}, j \in \{l_{k=1}\}}$ connecting the $l_k$ neurons of layer $k$ to the $l_{k+1}$ neurons of layer $k+1$. In each neuron an activation function is applied on the weighted sum of the inputs to compute the output.There are different activation functions, we want to mention some popular ones: \\
\begin{compactitem}
	\item Linear function: $f(x) = ax$
	\item Sigmoid function: $\sigma(x) = \dfrac{1}{1 + e^{-x}}$
	\item Tanh function: $tanh(x) = \dfrac{e^{2x} -1}{e^{2x} + 1}$
	\item ReLu\footnote{Rectified linear unit} function:  $f(x) = max(0,x)$ gives an output $x$ if $x$ is positive and $0$ otherwise 
	\item PReLu\footnote{Parametric ReLu} function: $f(x) = max(x, ax), a \leq 1$ with $a$ as parameter that will change depending on the model. 
	\item Softmax function: $\sigma(x)_j = \dfrac{e^{z_j}}{\sum_{k=1}^{K}e^{z_k}}$, usually used when we want to address the multiple class scenario with $k$ classes
\end{compactitem}
It is out of the scope of this work to further explain when to use which activation function or how to build the architecture of the network before training.
In this context we assume only \acp{nn} in supervised \ac{ml}. A common learning algorithm for \acp{nn} is again the gradient descent algorithm, with help of back propagation. Again, the aim is to minimize the cost-function. But instead of modifying just one weight vector, in a \ac{nn} we need to adapt all weights of each node of each layer. So in each iteration the algorithm first calculates the overall costs of the training data set, called forward pass or forward propagation, afterwards the algorithm, starting at the output layer, adapts the weights of the previous layer, called backward pass or backward propagation. The \sota learning algorithms in the \pp context are mostly optimizations of the gradient descent method. 
\subsection{k-Nearest-Neighbor ($k$NN)}
Although the $k$NN algorithm belongs to the supervised algorithms, it does not directly learn a model like the previous ones and therefore does not need a training phase before the evaluation phase. Rather it stores all the training data and classifies new data records based on similarity measures. $k$ is the previous defined constant and for evaluating the data record is assigned to the class which is most frequent among the $k$ nearest training samples. Because this algorithm requires to store the training data at the party that later wants to evaluate new data records, the $k$NN is not fit for our use case and we will not go into further detail.

\subsection{Distributed Machine Learning}
Since in the case of \ac{he} and \ac{tee} our use case requires to have the data partitioned so that no server has access to the whole data set, we need to consider distributed learning, also called data parallelism. Since the upcoming of big data the parallelism of \ac{ml} is a focus of research. 
A widely examined and adopted way is data parallelism in gradient descent based learning algorithms (e.g. linear regression, logistic regression, Neural Networks). The data is divided so that each server has one part of the data set and the model. Each server operates on its subset of the data and predicts the errors between the training samples and the labeled outputs. After that the server update their model based on the errors and need to communicate the changes to the other nodes. So in each iteration the servers need to synchronize the model parameters, or gradients, at the end of the batch computation to ensure they are training a consistent model.
The training of decision trees in a distributed way is not quite as obvious as in gradient descent, but again there are several algorithms that can be used, e.g. \cite{panda2009planet},  \cite{ben2010streaming}, \cite{Meng2016MLlib}, \cite{Si2017}.
Especially when dealing with big data without the \pp approach there exists frameworks that handle distributed \ac{ml} like Apache Spark \cite{Meng2016}.
	\section{Existing Solutions}\label{sec:exSolutions}

\subsection{Surveys on privacy-preserving ML}
Several paper give an overview of existing solutions in \pp \ac{ml}. We want to present some of them. %
\cite{CUNHA_Survey} consider different types of heterogeneous data and \pp mechanism that can be used for these data types. They consider not only cryptographic methods but their main focus lies on other methods like anonymization and obfuscation. With this work they aim to show which \pp mechanism are applicable for the characteristics of each data type.
\cite{Kanwal2021} review privacy-aware anonymity-based techniques for electronic health records for heterogeneous data types but do not consider cryptographic methods. 
\cite{Yang2019} provide a technical review and comparison of existing solutions for outsourced computation. They look at secure \ac{mpc}, pseudorandom functions, software guard extensions, perturbation approaches but mainly aim to provide an overview of the existing secure outsourcing solutions based on \ac{he} algorithms. A wide range of applications like regression training or biometric authentication is considered. They compare the security and efficiency performances.
\cite{Azraoui2020} provide a overview of existing \ac{he} and \ac{mpc} solutions for the problem of privacy preserving \acp{nn}, whereas \cite{sokClustering} focuses on clustering, a popular unsupervised machine learning algorithm and considers \ac{he} as well as \ac{mpc}. They claim to be the first survey that is concerned with an unsupervised \ac{ml} algorithm.
\cite{Chatel2021} performs a cross-field systematization of knowledge of privacy-preserving tree-based models. They consider differential privacy based solutions as well as cryptographic solutions like \ac{mpc} and \ac{he}.
\cite{Song2020} investigate the \sota works to resolve training of \ac{ml} in a \pp way. They consider secure \ac{mpc} and federated learning solutions. They categorize secure \ac{mpc} frameworks into \ac{he}-based,\ac{gc}-based and \ac{ss}-based and mixed learning frameworks.
As far as we know, we are the first that consider several \ac{ml} algorithms in the context of cryptographic solutions focusing on the client-server model where several clients provide their data in a \pp way and do not want to be involved with the computation.
Next we want to consider some recent \sota solutions using the explained cryptographic solutions, \ac{he}, \ac{mpc} and to some extend also \ac{tee}. We will categorize them by the used technique and the trained algorithm, considering only solutions that consider \pp \ac{ml} training in the medical context.

	\subsection{\ac{he} based Solutions} \label{subsec:HE}
In this section we want to introduce some \sota paper considering homomorphic \ac{ml}. Although our use case in the \ac{he} scenario designs a distributed learning, we will see only classic one server setups, since we could not find any paper regarding distributed homomorphic \ac{ml}. Table \ref{tab:tableHE} gives an overview over the paper considering \ac{he}.
\begin{table}
	\begin{scriptsize}
		\begin{center}
			
			\begin{tabular}{|l|cccc|cccc|l|l|}
		
		\multicolumn{1}{c}{}& \multicolumn{1}{c}{ \rot{linear Regression}} 	&\multicolumn{1}{c}{ \rot{logistic Regression}} &\multicolumn{1}{c}{ \rot{Decision Trees}} &\multicolumn{1}{c}{ \rot{NN}} & \multicolumn{1}{c}{\rot{CKKS}}  & \multicolumn{1}{c}{\rot{FV} }& \multicolumn{1}{c}{\rot{BGV}}  & \multicolumn{1}{c}{\rot{TFHE} } & \multicolumn{1}{c}{} &  \multicolumn{1}{c}{}\\ 		
			\multicolumn{1}{c}{\textbf{Paper}} 	& \multicolumn{4}{c}{\textbf{Algorithm}} & \multicolumn{4}{c}{\textbf{Scheme}} &\multicolumn{1}{c}{\textbf{Data Set$^\star$}} & \multicolumn{1}{c}{\textbf{Time}} \\ 
	
				\hline 	\hline 
				\cite{Giacomelli2018}$^*$ & \CIRCLE & \Circle & \Circle & \Circle & \Circle& \Circle& \Circle& \Circle & $10^4$ r, 20 f & 10.28 s \\ \hline
					\cite{Bonte2018}&\Circle   &  \CIRCLE & \Circle  & \Circle & \Circle &\CIRCLE & \Circle & \Circle & (7) 245 r, 3 f,  10643 SNPs  & 44 min \\
					\hline
				
					\multirow{2}*{\cite{Chen2018}}&	\multirow{2}*{\Circle}&   \multirow{2}*{\CIRCLE} & 		\multirow{2}*{\Circle}& 	\multirow{2}*{\Circle} &  	\multirow{2}*{\Circle} & 	\multirow{2}*{\CIRCLE} & 	\multirow{2}*{\Circle}	& 	\multirow{2}*{\Circle} & (5) 1579 r, 108 f & 14.9/115.33 h \\
					& & && &&& & & 1500 r, 196 f & 27.1/48.76 h \\ \hline
						\multirow{6}*{	\cite{Kim2018-2}}&	\multirow{6}*{\Circle}&  	\multirow{6}*{\CIRCLE} &  	\multirow{6}*{\Circle}& 	\multirow{6}*{\Circle} & 	\multirow{6}*{\CIRCLE} & 	\multirow{6}*{\Circle}	& 	\multirow{6}*{\Circle}& 	\multirow{6}*{\Circle} & (5) 1579 r, 108 f  & 8 min \\
							& & & && & 	&&&														(1) 1253 r, 10 f & 	3.6 min	\\
					& & & & & & &		&&														(2) 189 r, 10 f	 & 	3.3 min	\\
					& & & & & & 	&	&&														(3) 15,649 r, 16 f & 	7.3 min	\\
					& & & & & & 	&	&&													(4) 379 r, 10 f & 	3.5 min	\\
					& & & & & & 	&	&&											(6) 575 r, 9 f & 	3.5 min	\\	\hline
					\cite{Qiu2020}$^*$ & \CIRCLE & \Circle & \Circle & \Circle & \Circle& \Circle& \Circle& \Circle & $10^4$ r, 20 f & 64.35 s \\ \hline
				\multirow{5}*{	\cite{Kim2018}} & 	\multirow{5}*{\Circle} & 	\multirow{5}*{\CIRCLE} & 	\multirow{5}*{\Circle}	& 	\multirow{5}*{\Circle} & 	\multirow{5}*{\CIRCLE} & 	\multirow{5}*{\Circle}	& 	\multirow{5}*{\Circle}& 	\multirow{5}*{\Circle} & (1) 1253 r, 10 f & 131 min \\
				&	& & & & &  && & 																	(2) 189 r, 10 f	 & 	101 min	\\
				&	&& & &  & & & & 																(3) 15,649 r, 16 f & 	265 min	\\
				&	& & &	 & & & & & 															(4) 379 r, 10 f & 	119 min	\\
				&	& & 		& & & & & & 															(6) 575 r, 9 f & 	109 min	\\																	 
				\hline
					\cite{akavia2020} & \Circle  & \Circle & \CIRCLE & \Circle &  \CIRCLE & \Circle & \Circle &\Circle & - & 19 - 1220 min \\ \hline
					 	\cite{Carpov2020} & \Circle & \CIRCLE & \Circle  & \Circle &  \CIRCLE &   \Circle & \Circle &  \CIRCLE & (7) 245 r, 3 f, 10643 SNPs  & 186 min \\
					 	\hline
					 	\cite{Lou2019} & \Circle & \Circle & \Circle & \CIRCLE & \Circle &\Circle & \CIRCLE& \CIRCLE & (10) 70000 r, 784 f & 8 days \\ \hline
			\end{tabular} \caption{State of the art homomorphic encryption based solutions. \\
		$^*$Using Pailier as partial \ac{he} scheme and data masking. \\$^\star$More information about the data sets can be found in the Appendix \ref{DataSets}.}\label{tab:tableHE}
	\end{center}\end{scriptsize}

\end{table}
\paragraph{Linear Regression \newline}
\cite{Giacomelli2018} addresses \pp linear regression via a partial homomorphic encryption scheme (e.g. Pailier). They assume several data owners, that want to share their information but only if it is encrypted, a Machine-Learning Engine (MLE) and a Crypto Service Provider (CSP). The MLE wants to run a linear regression algorithm on the merged data. The CSP generates a public/private key pair and sends the public key to data holders and the MLE. The data owner send their encrypted records to the MLE which performs homomorphically a masking algorithm and sends the encrypted masked data to the CSP. The CSP decrypts the masked data and performs the linear equation of the regression problem and sends the result back to the MLE which then demasks the result and obtains the real model.
\cite{Qiu2020} is very similar to \cite{Giacomelli2018}, but they use a different masking technique. 
The papers presented above both assume a similar use case as we do and the combination of homomorphic encryption and data masking realizes a protocol ideally suited for practical applications. One drawback is, that the use of the CSP introduces communication costs, that are prevented in the standard \ac{he} scenario.
\paragraph{Logistic Regression \newline }
\cite{Kim2018} train the logistic regression with the HEAAN scheme \cite{HEAAN} using gradient descent. Since the sigmoid function cannot be directly evaluated with an \ac{he} scheme it needs to be approximated. They use a global approximation method for the sigmoid function of the logistic regression minimizing the mean squared error. They test their implementation with degree 3 and 7 least square approximation while degree 3 requires a smaller depth for evaluation and degree 7 polynomial has better precision. To speed up the computation they use a packing mechanism to perform several evaluations in parallel. Instead of running the gradient descent until convergence, they use a fixed number of iterations. The runtime of their implementation needs between 265 and 94 minutes depending on the used data set.
\cite{Chen2018} are participants of the \ac{idash}\footnote{Since 2014 the \ac{idash} competition is yearly hosted. The competition in the biomedical sector is concerned with privacy enhancing technologies around the theme of genomic and biomedical privacy. They aim to bring experts in security, privacy and bioinformatics together to test the limits of secure computations.} 2017 competition. In the year 2017 one of the tasks was to train a logistic regression model over encrypted genomic data to predict disease based on a patients genome. They used the FV-scheme \cite{fvScheme} with batching and modified the gradient decent method to minimize the increase in the size of the numbers. For this they designed two different variants. The first variant named 1-Bit Gradient Descent updates in each iteration the weight by a learning rate multiplied by the sign of the current gradient, in the second they use the original gradient descent with an approximated sigmoid function. Additionally they modified the bootstrapping method to combine boot with scaling to prevent plaintext size expansion. They tested the both algorithms on the \ac{idash} dataset with 36 iterations of the Gradient Descent and also on a modified MNIST dataset, where they only used 1500 images, containing the handwritten numbers 3" or "8" and reduced every image to 196 features. They did 10 iterations on the MNIST dataset.
\cite{Bonte2018} also participated in the \ac{idash} 2017 competition and also used the FV-scheme. They used in contrast to \cite{Chen2018} not the gradient descent but rather applied a new iterative method that uses a simplified fixed hessian method to reduce the multiplicative depth. In the practical training they only did one iteration with a fixed number of covariates (=20) and needed 44min for training on 1000 records. 
\cite{Kim2018-2} are the winners of the task of the \ac{idash} competition 2017, participating with a implementation based on the HEAAN-scheme  \cite{HEAAN}.  They used a new encoding method which reduces the required storage space and optimizes the computational time. They adapted the Nesterov’s accelerated gradient descent algorithm to reduce the needed iterations and therefore increase the speed. 
\cite{Carpov2020} are one of the finalist of the \ac{idash} 2018 competition, where the task was to find all features in an encrypted data base that will improve the quality of an already trained regression model. Concretely the data analyst builds a logistic regression model using phenotype features and updates this model afterwards with genotype features. The training of the logistic regression in Step 1 uses TFHE \cite{TFHE}, the updating process is build with HEAAN \cite{HEAAN}.
\paragraph{Decision Trees \newline}
\cite{akavia2020} propose a new \pp solution for training and prediction of tree-based algorithms. As a central technique they developed a soft-step function yielding in a low-degree approximation (degree) and a lightweight interactive protocol, that is independent of the tree or data set size. They evaluated several UCI data sets on a single decision tree using \cite{HEAAN} with the SEAL-library and achieved a runtime between 19 minutes (100 examples, 4 features, 3 labels) to 1220 min (10000 examples, 54 features, 7 labels) depending of the data set size.
\paragraph{Neural Networks \newline}
\cite{nandakumar2019towards} are the first to address the training of a \ac{nn} in encrypted form using the BGV scheme. The architecture of the \ac{nn} used in this work is a 3-layer fully connected network with sigmoid activation function that are realized with look up tables. They need for one mini-batch containing 60 training samples about 40 min. Because they do not provide the full training costs this work is not presented in Table 3.  
\cite{Lou2019} propose Glyph, a \ac{fhe}-based technique for \ac{nn}. They use TFHE for implementing activation functions such as Relu and softmax and switching to vectorial-arithmetic-friendly BGV when processing fully-connected and convolutional layers. They test their implementation on various data sets for example the MNIST data set.

	\subsection{\ac{mpc} based Solutions} \label{subsec:MPC}
Since \ac{mpc} is an older technique than \ac{he} it is already widely used in practice and there are more paper regarding \ac{mpc}. Additionally the complexity of algorithms considered in the \sota solutions has increased. While in \ac{he} barely one paper considers \ac{nn}, in \ac{mpc} deep \acp{nn} are the main focus of recent research. Therefore we will mainly consider \acp{nn} in this Chapter and will only present a few paper to other machine learning algorithms.
As stated in Section \ref{sec:techniques} we will only consider papers that accomplish our requirement of a distributed data set among the servers via secret sharing. Table \ref{tab:tableMPC} gives an overview over the paper considering \ac{mpc}.
\begin{landscape}

\begin{table}
	\begin{scriptsize}
		\begin{center}
		
			\begin{tabular}{|l|c|cccc|ccc|cc|l|l|l|}
				
				\multicolumn{1}{c}{}	&	\multicolumn{1}{c}{}	&\multicolumn{1}{c}{ \rot{linear Regression}} 	&\multicolumn{1}{c}{ \rot{logistic Regression}} &\multicolumn{1}{c}{ \rot{Decision Trees}} & \multicolumn{1}{c}{ \rot{NN}} & \multicolumn{1}{c}{\rot{2PC}}  & \multicolumn{1}{c}{\rot{3PC} }& \multicolumn{1}{c}{\rot{4PC}}  & \multicolumn{1}{c}{\rot{passive} }& \multicolumn{1}{c}{\rot{active} } & \multicolumn{1}{c}{} &  \multicolumn{1}{c}{} &  \multicolumn{1}{c}{}\\ 	
				\multicolumn{1}{c}{\textbf{Paper}} & \multicolumn{1}{c}{\textbf{Framework}} 	& \multicolumn{4}{c}{\textbf{Algorithm}} & \multicolumn{3}{c}{\textbf{Parties}}&  \multicolumn{2}{c}{\textbf{sec. model}} &\multicolumn{1}{c}{\textbf{Data Set}} & \multicolumn{1}{c}{\textbf{Time}} &\multicolumn{1}{c}{\textbf{Comm. Cost}} \\ 
				\hline
				\hline 
			\multirow{3}*{$\cite{mohassel2017}^\ast$}& \multirow{3}*{SecureML} & \CIRCLE & \Circle & \Circle  & \Circle & \multirow{3}*{\CIRCLE} & \multirow{3}*{\Circle}  & \multirow{3}*{\Circle} & \multirow{3}*{\CIRCLE}  & \multirow{3}*{\Circle}  &\multirow{3}*{ $10^6$  r, 784 f} & 50 s - 100 s & \multirow{3}*{-}  \\
		&	& \Circle & \CIRCLE & \Circle & \Circle & & && & &   & 100 s - 150 s  &  \\ 
			&	& \Circle  & \Circle & \Circle & \CIRCLE && & & & &   & 653 s - 4239.7 s   &  \\ 	\hline
						\multirow{2}*{	\cite{Shi2016}}& 	\multirow{2}*{SMAC-GLORE} & 	\multirow{2}*{\Circle} & 	\multirow{2}*{\CIRCLE} & 	\multirow{2}*{\Circle}  & 	\multirow{2}*{\Circle} & \CIRCLE& \Circle& \Circle & 	\multirow{2}*{\CIRCLE}  & 	\multirow{2}*{\Circle} & 	\multirow{2}*{60 r, 3 f} &  7290.67 s & - \\ 
						& & & & & & \Circle & \CIRCLE& \Circle&& & & 17862.47 s & \\ \hline
				\cite{Chen2019}& SPDZ &\Circle & \CIRCLE & \Circle  & \Circle & & &&  \CIRCLE & \Circle & 11 & 16.45 s & - \\ \hline
				\multirow{2}*{\cite{deCock2021}$^\star$}& \multirow{2}*{-} & \multirow{2}*{\Circle} & \multirow{2}*{\CIRCLE} & \multirow{2}*{\Circle} & \multirow{2}*{\Circle}  &\multirow{2}*{\Circle} & \multirow{2}*{\LEFTcircle} &\multirow{2}*{\Circle} & \multirow{2}*{\CIRCLE}  & \multirow{2}*{\Circle} & (8) 470 r, 17814 f  & 2.52 s & - \\
				&	&&&&&&&&&& (9) 225 r, 12634 f & 26.9 s & - \\ \hline
				\cite{De2014}& VIFF & \Circle & \Circle & \CIRCLE & \Circle &\Circle & \CIRCLE&\Circle &\CIRCLE  &\Circle & 3196 r & 73 s & - \\ \hline
				\cite{Adams2021}& - & \Circle & \Circle & \CIRCLE & \Circle &\CIRCLE &\Circle &\Circle & \CIRCLE & \Circle & (9) 225 r, 12634 f & 12.6 s & - \\ \hline
				\multirow{2}*{	\cite{Abspoel2021}}&\multirow{2}*{MP-SPDZ}& \multirow{2}*{\Circle}	 & \multirow{2}*{\Circle} & \multirow{2}*{\CIRCLE} & \multirow{2}*{\Circle} & \multirow{2}*{\Circle} & \multirow{2}*{\CIRCLE} &\multirow{2}*{\Circle} &  \CIRCLE & \Circle & \multirow{2}*{8192 r} &  34 s & 3.783 GB  \\ 
				&&  &  & & &  &&  & \Circle  &  \CIRCLE&  &  182.1 s & 16.552 GB \\ \hline
				\cite{Wagh2018} & SecureNN &\Circle & \Circle  & \Circle & \CIRCLE & \Circle & \CIRCLE &  \Circle & \CIRCLE &  \Circle & (10) 70000 r, 784 f & 0.88 h & 113 GB \\ \hline
				\cite{Mohassel2018}& Aby3 &\Circle & \Circle  & \Circle & \CIRCLE & \Circle & \CIRCLE &  \Circle & \CIRCLE &  \Circle &(10) 70000 r, 784 f & 0.75 h & 31.744 GB\\ \hline
			\cite{Agrawal2019}& QOUTIENT &\Circle & \Circle  & \Circle & \CIRCLE & \CIRCLE  & \Circle&  \Circle & \CIRCLE & \Circle & (10) 70000 r, 784 f & 50.25 h & - \\ \hline
				\multirow{2}*{	\cite{Wagh2020}}& \multirow{2}*{FALCON} & \multirow{2}*{\Circle}	 & \multirow{2}*{\Circle} & \multirow{2}*{\Circle} & \multirow{2}*{\CIRCLE} & \multirow{2}*{\Circle} & \multirow{2}*{\CIRCLE} &\multirow{2}*{\Circle} &  \CIRCLE & \Circle & \multirow{2}*{(10) 70000 r, 784 f} &  0.17 h & 16.384 GB  \\ 
			&	&  &  & & &&  &  & \Circle  &  \CIRCLE&  &  0.56 h & 90.112 GB \\ \hline
					\multirow{2}*{	\cite{Attrapadung2021}}& \multirow{2}*{-}  & \multirow{2}*{\Circle}	 & \multirow{2}*{\Circle} & \multirow{2}*{\Circle} & \multirow{2}*{\CIRCLE} &   \multirow{2}*{\Circle} & \multirow{2}*{\CIRCLE}  &\multirow{2}*{\Circle}  &  \CIRCLE & \Circle & \multirow{2}*{(10) 70000 r, 784 f} &  117 s &  \multirow{2}*{-} \\ 
			&	&  &  & & &  &&  & \Circle  &  \CIRCLE&  &  570 s &  \\ \hline
				\cite{Ge2021} & - & \Circle  & \Circle  & \Circle & \CIRCLE & \CIRCLE  & \Circle&  \Circle & \CIRCLE & \Circle & (10) 70000 r, 784 f & 0.238 h & 48.128 GB \\ \hline
			\end{tabular} \caption{State of the art multi-party computation based solution. The runtimes are always in the LAN-Setting.\\$^\ast$Considering only the online phase for the running time. \\$^\star$ One party who pre-computes multiplication triplets, two parties that are actively computing the final result.}\label{tab:tableMPC}
	\end{center}\end{scriptsize}	
\end{table}
\end{landscape}

\paragraph{Linear Regression \newline}
\cite{mohassel2017} present a new and efficient protocol for \pp machine learning not only for linear regression but also for logistic regression and neural networks. They were the first to present a \pp protocol for the latter two. The setup is the two-server model where data owners distribute their private data among two non colluding servers that train via secure two-party computation. To improve the runtime they propose \ac{mpc} friendly alternatives to the non-linear functions (e.g. sigmoid function) and develop new techniques to support secure arithmetic operations on shared decimal numbers. Their protocol has an offline and a online phase. In the offline phase they generate shared multiplication triplets and present two protocols, one with linearly homomorphic encryption (LHE) and the other with \ac{ot}.
\paragraph{Logistic Regression \newline} 
\cite{Shi2016} propose a  \ac{ss} , circuit-based secure \ac{mpc} framework (SMAC-GLORE) considering the client-server setup, that not only protects patient-level data but also all intermediary information. They implement the Newton-Raphson method for learning on the data. 
\cite{Chen2019} investigates the efficiency of the SPDZ framework \cite{Damgaard2012} that provides malicious security. The compare their results to applications implemented with semi-honest \ac{mpc} techniques and show that they easily outperform the previous implementations while providing stronger security. 
\cite{deCock2021}  train a logistic regression model via a secure two-party protocol using a trusted initializer in an offline phase. The initializer distributes correlated randomness (multiplication triples) and can operate in an offline phase before the data is known. The other two parties compute the final result in an online phase using additive \ac{ss}. They implement a new protocol for the activation function, that does neither require secure comparison protocol nor Yao's garbled circuits. This paper is an improvement of one of the winners in track 4 of the \ac{idash} 2019 competition. Unfortunately they do not present the communication costs. As above described \cite{mohassel2017} also present a protocol for logistic regression.
\paragraph{Decision Trees \newline} 
\cite{De2014} consider the ID3 algorithm for decision tree learning via \ac{mpc} using \ac{ss}. Their evaluation needs only a few seconds to minutes, depending on the used data set. 
\cite{Adams2021}  address the problem of \ac{mpc} unfriendly learning algorithms. They observed that the standard algorithm for training decision trees with continuous features (C4.5 algorithm) that is used for cleartext data requires sorting of training examples, which is very expensive in \ac{mpc}. Therefore they propose three different, more efficient alternatives. Additionally they propose several improvements and optimizations to important building blocks of \pp \ac{ml} protocols. Training with the alternatives needs a few minutes and achieves accuracy that is the same with those obtained in the clear. With the new algorithms they present the first protocols that can handle continuous data without relying on a full sorting of the data set.
In contrast \cite{Abspoel2021} too consider training of decision trees on continuous data but stick with the sorting of the data using a modified and stripped down version of C4.5. They achieve a protocol that can handle both discrete and continuous data and also both semi-honest (passive) and malicious (active) adversary. They implement their protocols using the MP-SPDZ framework \cite{mp-spdz}.
\paragraph{Neural Networks \newline}
Although MNIST is not a medical data set, but rather contains handwritten digits we will present the runtime of the considered paper with the MNIST dataset because it is typically used as a benchmark in the context of \acp{nn}. \newline
As above described \cite{mohassel2017} also present a protocol for \acp{nn}. %
\cite{Mohassel2018} design and implement a new framework for \pp \ac{ml} in a three-server model with a single corrupted server, called ABY3. Beside a new approximate fixed-point multiplication protocol they also develop a new protocol for efficiently converting between binary sharing, arithmetic sharing and Yao sharing. Unfortunately they do not state their evaluation run time for the training of the \ac{nn} in their maliciously setting but only for the semi-honest threat model.
\cite{Wagh2018} developed new information-theoretically secure protocols for various building blocks of neural networks and therefore achieve very efficient protocols for training of different \ac{nn} architectures. They test their system called SecureNN on the MNIST Data Set and need for a single image inference 0.04s and 2.08MB communication. In contrast to \cite{mohassel2017} they achieve a speed up 98x and 8x for the 2 and 3-party setting.
QUOTIENT\cite{Agrawal2019} is a new framework for discretized training of deep \ac{nn} in the 2 party setting. They do not need an offline phase and test their implementation on various data sets yielding in a performance for the MNIST data set with a 3 x (128FC) \ac{nn} and 5 training epochs of 50.25 h.
\cite{Wagh2020}	supports a new protocol in a honest-majority 3-party setting called FALCON. They improve the round and communication complexity and provide security against maliciously corrupt adversaries with an hones majority. Additionally they are the first that demonstrate efficient protocols for batch-normalization.
\cite{Attrapadung2021} propose ADAM a new secure and efficient protocol for a set of elementary functions that are useful for depp \acp{nn}, including secure division, exponentiation, inversion and square root extraction. They aim to use state of the art \ac{mpc}-unfriendly algorithms without approximation using \ac{mpc}-friendly functions, for example the Adam algorithm. 
\cite{Ge2021} implement a 2PC protocol to train a \ac{nn}. They consider only the passive - semi-honest security model. They build a new preprocessing protocol for mask generation, support and realize secret sharing comparison and aim to further reduce the communication costs. 
%
	
	\subsection{TEE based Solutions} \label{subsec:TEE}

Since neither the computational effort nor the communication costs play a big role in \ac{tee}, the task of training or evaluating a machine learning algorithm is not necessarily a challenge. Therefore there exists only  few paper in the recent research concerning simple machine learning tasks \cite{Leung2019towards}, \cite{law2020secure}. Rather there is a lot of work to prevent solutions with \ac{tee} for the standard attacks on \ac{tee} like side channel attacks  \cite{law2020secure}. Another area is the connection of \ac{tee} and \ac{he} or \ac{mpc} \cite{fischer2017computation}, \cite{jiang2018securelr}, which we will not explore further in this context.

	\section{Conclusion}

We discussed a lot of implemented solutions that are possible in our unique use case. Although we have presented runtimes for most of the implementations, these can only be seen as a rough guideline, as the implementations differ in several important details. On the one hand, the computing power of the devices used is often very different, on the other hand, several data sets are used for bench marking, and finally, the number of iterations of the training algorithm is not handled uniformly.
Most of the \ac{ml} algorithms are possible in a distributed way, for example in the \ac{he} scenario, especially those that are trained via the gradient descent method. The logistic regression has various advantages in the medical field. It is easy to train and already widely used and it is easy to interpret the trained model, which is useful in the medical context. \acp{nn} on the other hand can detect more complex relations but are costlier to train and not really interpretable. Decision Trees are easy to understand and are therefore also often used but they are not as suitable for our distributed context and again costlier to train as the logistic regression. But the choice of the algorithms naturally depends on the research question and the existing data. 
If we look at the cryptographic methods that can be used in our use case we can see the standard advantages and disadvantages. \ac{tee} are efficient and also common but they bring along a trust issue, since the data will be decrypted in the enclave. \ac{he} usually is used with one \ac{dh} and one \ac{dp} as outsourced computation. A big advantage of \ac{he} over \ac{mpc} is that there are no communication costs. On the other hand the computations are still slower than in \ac{mpc}. 
If we assume a distributed training we need to introduce additional communication costs in both scenarios, \ac{tee} and \ac{he}, which takes away the advantage. In \ac{he} we do not need to protect the communication between the servers, because all data and intermediate results are encrypted anyways, in \ac{tee} there should be secure communication channels.
Then again, a disadvantage of \ac{mpc} in our scenario with shared data is that we need robust computing server. If one computer fails the computation cannot be done in the most frameworks. In the scenario with \ac{tee} or \ac{he} the data is distributed in a horizontal way (ore sometimes also an vertical way) where failing of a server often only means a smaller data set for training.

\newpage	

	\bibliographystyle{plain}
	\bibliography{report}
	\newpage
\appendix
\section{Data Sets} \label{DataSets}
There are a variety of data sets from fully synthetic to partially synthetic to real data. Sometimes there is a reference data set for certain algorithms, such as the MNIST for \ac{nn}, but often the research paper use some synthetic data to test the performance of their solutions with a set number of records and features. Whenever they use a special and published data set, we listed it in Table \ref{tab:Dataset}.
\newline
\begin{table}[h]
	\begin{center}
		\label{tab:Dataset}
		\begin{tabular}{|cl|c|c|}
			\hline
			\textbf{} &	\textbf{Data Set}  & \textbf{records} & \textbf{features} \\
			\hline
			1  &	Edinburgh Myocardial Infarction \cite{kennedy1996early} & 1253 & 10  \\  \hline 
			2&	Low Birth Weight Study \cite{LBW} & 189 & 10\\  \hline
			3&	Nhanes III \cite{Nhanes} & 15,649 & 16\\  \hline
			4&	Prostate cancer Study \cite{Prostata} & 379 & 10\\  \hline
			5&		\ac{idash} 2017 & 1579 & 108 \\ \hline
			6&	Umaru Impact Study \cite{Umaru} & 575 & 9 \\  \hline
			7&	\ac{idash} 2018 & 245 & 3\\ 
			&	& 	& 10643 SNPs \\\hline
			8&	BC-TCGA (\ac{idash} 2019-1)  & 470 & 17814 \\ \hline
			9&	GSE2034 	(\ac{idash} 2019-2)  & 225 & 12634   \\ \hline
			10&			MNIST \cite{MNIST} & 70000 & 28x28 pixel = 784 \\ \hline
			11 & CT slices \cite{CTSlices} & 53 500 & 384  \\ \hline
		\end{tabular}\caption{Overview of data sets}	
	\end{center}
\end{table}
	
\end{document}